\documentclass[aps,preprint]{revtex4}
\usepackage{amssymb}
\usepackage{amsmath}
\usepackage{graphicx}
\usepackage{epstopdf}

\begin{document}

\title{\bf The effect of a density dependent bag constant on the structure of  hot neutron star with a quark core}

\author{{\bf T. Yazdizadeh$^1$ and G. H. Bordbar$^{2,3}$}}
 \affiliation{
$^1$Islamic Azad University, Bafgh Branch, Bafgh, Iran\\
 $^2$Department of
Physics, Shiraz University,
Shiraz 71454, Iran (e-mail: bordbar@susc.ac.ir)\\
$^3$Research Institute for Astronomy and Astrophysics of Maragha,\\
P.O. Box 55134-441, Maragha 55177-36698, Iran }

\begin{abstract}
As we go from center toward the surface of a neutron
star, the state of baryonic matter changes from the de-confined
quark-gluon to a mixed phase of quark and hadronic matter, and a
thin crust of hadronic matter. For the quark matter, within MIT bag
model, the total energy density of the system is the kinetic energy
for non-interacting quarks plus a bag constant. In this article
first we have considered a density dependent bag constant obtained
using the recent experimental results of CERN SPS on the formation
of a quark-gluon plasma. For calculations of the hadron phase, we
use the lowest order constrained variational method. The equation of
state of mixed phase has been determined using Gibbs conditions.
Finally, we have calculated the structure of a hot neutron star with
quark core employing TOV equation. Our results show that a density
dependent bag constant leads to a higher mass and lower radius for
the hot neutron star with respect to the case in which we use a
fixed bag constant.
\end{abstract}
\maketitle
\section{Introduction}\label{S:intro}
Neutron stars (NS) are among the densest of massive objects in the
universe. A hot neutron star is born following the gravitational
collapse of the core of a massive star just after the supernova
explosion. The interior temperature of a neutron star at its birth
is of the order $20 - 50\ MeV$
\cite{burrows1986}. As we go from the surface to the center of a
neutron star, at sufficiently high densities, the matter is expected
to undergo a transition from hadronic matter where the quarks are
confined inside the hadrons to a state of deconfined quarks, with
up, down and strange quarks in the quark matter. Other quarks have
higher masses and do not appear in this state. Glendenning has shown
that a proper construction of the hadron-quark phase transition
inside the neutron stars implies the coexistence of nucleonic matter
and quark matter over a finite range of pressure. Therefore, a mixed
hadron-quark phase exists in the neutron star where its energy is
lower than that of the quark matter and the nucleonic matter over a
finite rang of Pressure \cite{glendenning1992}. This
shows that we can consider a neutron star to be composed of a
hadronic matter layer, a mixed phase of quarks and hadrons and, in
the core, quark matter.

The deconfined quark phase is treated within the popular MIT bag
model \cite{chodos1974}. In this model, the total
energy density is the sum of a nonperturbative energy shift ${\cal
B}$ (the bag constant) and the kinetic energy for noninteracting
quarks. The bag constant ${\cal B}$ can be interpreted as the
difference between the energy densities of the perturbative vacuum and physical ones, which has a constant value such as ${\cal B} =
55$ and $90 MeV/fm^3$ in the initial model of MIT, recently it is
constrained to be compatible with the recent experimental results
obtained at CERN on the formation of a quark-gluon plasma. The
resulting picture is the following: during the early stages of the
heavy-ion collision, a very hot and dense state (fireball) is formed
whose energy materializes in the form of quarks and gluons strongly
interacting with each other, exhibiting features consistent with
expectations from a plasma of deconfined quarks and gluons \cite{rafelsky1982, matsui1986}.
In general, it is not obvious
if the informations on the nuclear EOS from high energy heavy ion collisions
can be related to the physics of neutron stars interior. The possible
quark-gluon plasma produced in heavy ion collision is expected to be characterized
by small baryon density and high temperature, while the possible
quark phase in neutron stars appears at high baryon density and low temperature.
However, if one adopts for the hadronic phase a non-interacting
gas model of nucleons, antinucleons and pions, the original MIT bag model
predicts that the deconfined phase occurs at an almost constant value of the
quark-gluon energy density, irrespective of the thermodynamical conditions
of the system \cite{cleman1986}.

Burgio et al. have investigated the structure of neutron stars with
a quark core at zero \cite{Burgio2002a} and finite
temperatures \cite{burgio2007}, using the
Brueckner-Bethe-Goldstone formalism to determine the equation of
state of the hadronic matter. We have calculated the structure
properties of the cold neutron star by considering a quark phase at
its core \cite{bordbar2006} and compared the
results with our previous calculations for the neutron star without
the quark core \cite{bordbarh2006}. In these
works, we have employed the lowest order constrained variational
(LOCV) method for the hadronic matter calculations. Recently,  we
have calculated the structure of a hot neutron star with a quark
core with a fixed bag constant (${\cal B}=90\ MeV/fm^3$) \cite{yazdizadeh2011}. In the present paper, we intend
to extend these calculations to hot neutron stars with a quark cores
by considering a density dependent bag constant.


\section{ Equation of State }
\label{Energy calculation }
In this section, we calculate the equation of state of a
neutron star composed of a hadronic matter, a mixed phase of quarks and
hadrons and a quark core as follows.

\subsection{ Hadron Phase }
\label{Hadron phase}

In our calculations, the equation of state of hot nucleonic matter
is computed using the lowest order constrained variational (LOCV)
method \cite{bordbar1997, bordbar1998, modarres1998, bordbar2007a, bordbar2007b,
bordbar2008a, bordbar2008b, bigdeli2009}. The details of our calculations for the hadronic phase has been fully discussed in \cite{yazdizadeh2011}.


\subsection{Quark Phase}
\label{Quark Phase} We use the MIT bag model for the quark matter.
In this model, the energy density is the sum of kinetic energy of
quarks and a bag constant (${\cal B}$) which is interpreted as the
difference between energy densities of non interacting quarks and
interacting ones \cite{farhi1984},

\begin{equation}\label{eq6}
    {\cal E}_{tot} = {\cal E}_u + {\cal E}_d + {\cal E}_s + {\cal B},
\end{equation}
where ${\cal E}_i$ is the kinetic energy per volume of particle
$i$,
the kinetic energy of quarks has been discussed in \cite{yazdizadeh2011} and
The bag constant ${\cal B}$, can be interpreted as the difference
between the energy densities of the noninteracting quarks and
interacting ones, which has constant values such as ${\cal B} = 55,\
90$  and $220\ MeV$ in the initial model of MIT. But the density of
quark matter is not uniform, therefor we can consider a density
dependent ${\cal B}$. We try to determine a range of possible values
for ${\cal B}$ by exploiting the experimental data obtained at the
CERN SPS, where several experiments using high-energy beams of Pb
nuclei reported (indirect) evidence for the formation of a
quark-gluon plasma \cite{jacob2000, heinz2000}.
According to the analysis of those experiments, the quark-hadron
transition takes place at about seven times normal nuclear matter
energy density $({\cal E}_0 = 156 MeV fm^{-3})$. In the literature,
there are attempts to understand the density dependence of ${\cal
B}$ \cite{adami1993, blaschke1999}. However, currently the results are highly
model dependent and no definite picture has come out yet. Therefore,
we attempt to provide effective parametrization for this density
dependence. Our parametrization are constructed in such a way that
at asymptotic densities ${\cal B}$ approaches a finite value ${\cal
B}_{\infty}$. For the bag constant (${\cal B}$), we use a density
dependent Gaussian parametrization \cite{Burgio2002a, Baldo2006},
\begin{equation}\label{4}
{\cal B}(n)={\cal B}_{\infty}+\left({\cal B}_{0}-{\cal B}_{\infty}\right)
\exp[-\beta(n/n_{0})^{2}]
\end{equation}
with ${\cal B}_{0}={\cal B}(n=0)=400\ MeV/fm^{3}$ and
$\beta=0.17$.
We know that the value of the bag constant (${\cal B}$) should be
compatible with empirical results. The experimental results at
CERN-SPS show a proton fraction $ x_p = 0.4$ \cite{Burgio2002b}. Therefore, for calculation of ${\cal
B}_{\infty}$, we employ the equation of state of the asymmetric
nuclear matter as follows.
First, we use the equation of state of asymmetric hadronic matter
characterized by a proton fraction $ x_p = 0.4$ and the UV14 + TNI
potential. By assuming that the hadron-quark transition takes place
at the energy density ${\cal E} = 1100 MeV fm^{-3}$, we find that
the hadronic matter baryonic density is $ n_t $ (transition density)
and at values lower than $n_t$, the quark matter energy density is
higher than that of nuclear matter, while by increasing the baryonic
density, the two energy densities become equal at this density, and
after that, the nuclear matter energy density remains always higher.
Energy density equation of quark matter with two flavors u and d
reduces to
\begin{equation}\label{eq}
{\cal E}_Q={\cal E}_u+{\cal E}_d+{\cal B}
\end{equation}
Beta-equilibrium and charge neutrality conditions lead to the
following relation for the number density of quarks,
\begin{equation}\label{5}
n_B=2n_{u}=n_{d}.
\end{equation}
We determine ${\cal B}_{\infty}$ by putting quark energy density and
hadronic energy density equal to each other at any temperature.
Later we can calculate the energy density and determine pressure and
the equation of state for quark phase such as determined in \cite{yazdizadeh2011}.

\subsection{  Mixed phase   }
\label{Mixed phase} For mixed phase, where the fraction of space
occupied by quark matter smoothly increase from zero to unity, we
have a mixture of hadrons, quarks and electrons. According to the
Gibss equilibrium condition, the temperatures, pressures and
chemical potentials of both phases are equal \cite{glendenning1992}.
Calculation of  equation of state of mixed phase has been fully discussed in \cite{yazdizadeh2011}.

\subsection{Results}
\label{Results} 
\begin{figure}
\includegraphics[scale=0.45]{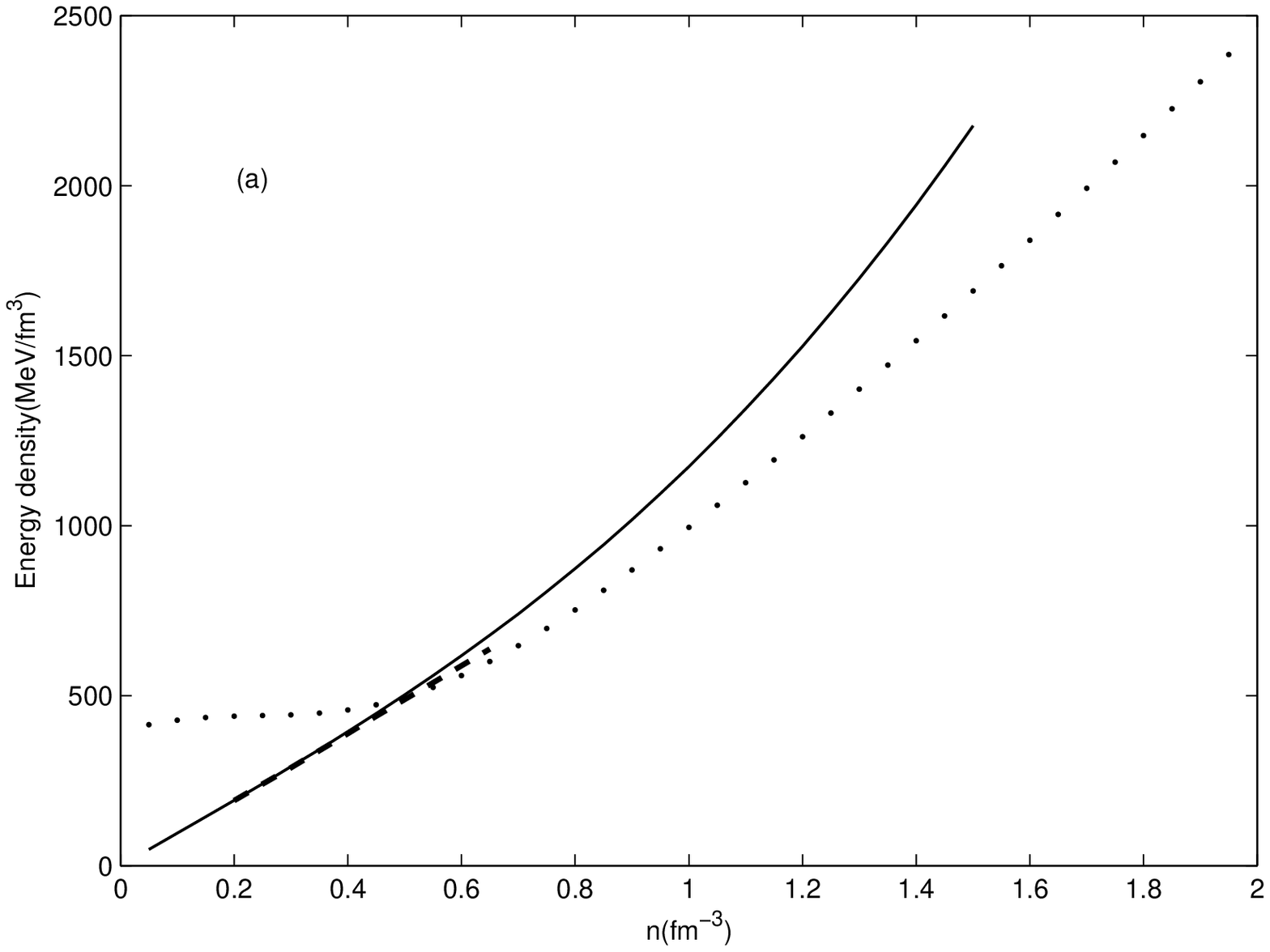}
\includegraphics[scale=0.45]{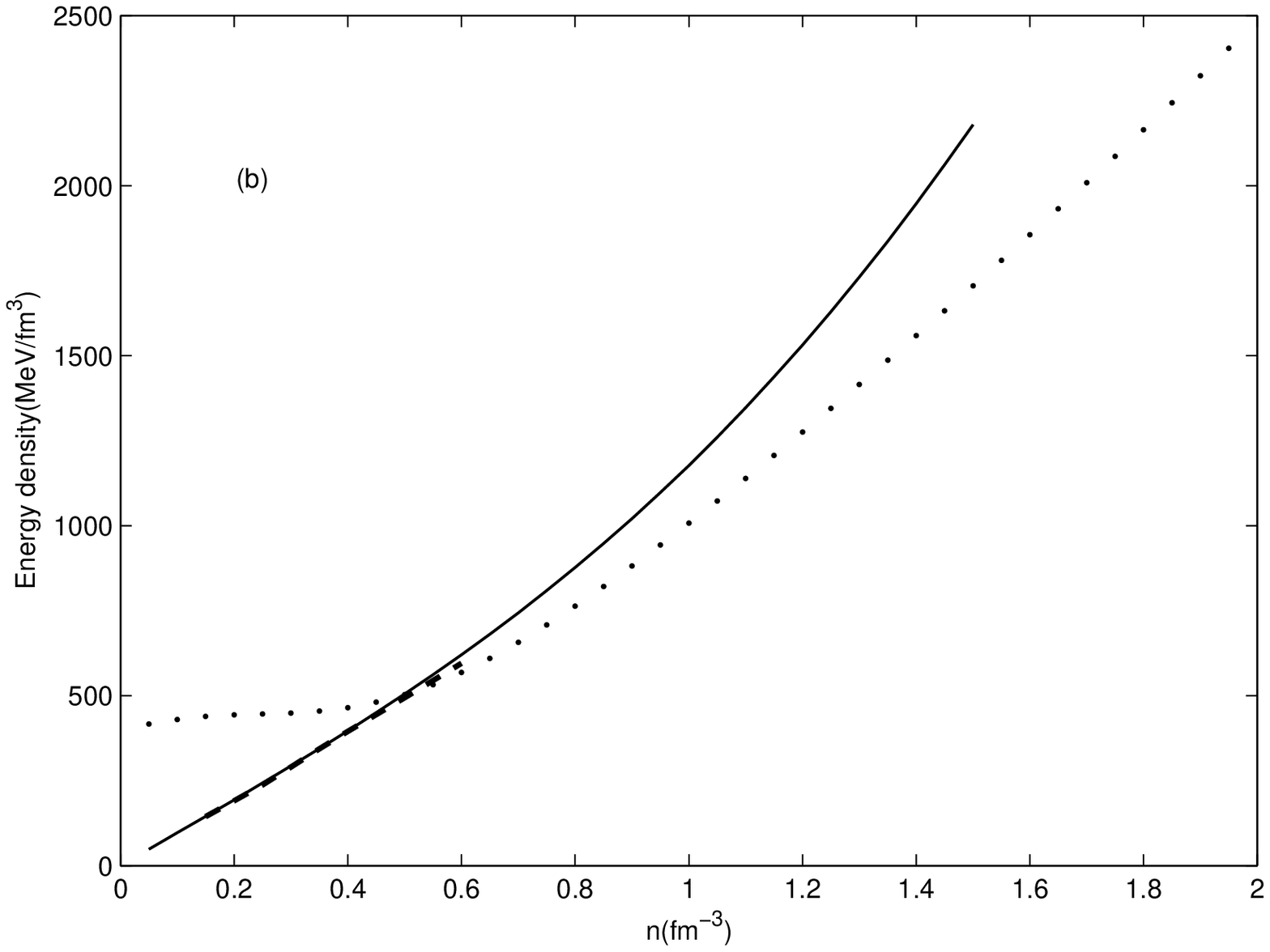}

\caption{Energy density versus the baryon density for the hadron phase
(solid line), mixed phase (dashed line) and quark phase (dotted line) at $T = 10$ (a) and $20\ MeV$ (b).}
\label{en1hmq}
\end{figure}

Our results for the energy densities corresponding
to different phases i.e.  hadronic matter, pure quark matter and a
mixed phase of quarks and hadrons, are given in Fig. \ref{en1hmq}
at two temperatures.
It can be seen that at low densities there is a pure hadronic phase.
We have found that a mixed phase exists at higher densities up to $n
\sim 0.5 fm^{-3}$. It is obvious that a pure quark phase emerges by
increasing the density.
The energy density for neutron star with density dependent and
density independent ${\cal B}$ are shown in Fig. \ref {en1f,n}.
It is clear that for a density dependent bag constant, the energy
density is smaller than that of the independent one.
We have calculated the equation of state (the pressure versus baryon
density) for the neutron star with the quark core using the density
dependent and independent bag constants. Fig. \ref {pn1f,n} shows
the relevant results at two temperatures. These
equations of state are used as an input into the general
relativistic equation of hydrostatic equilibrium.

\section{Structure of the Hot Neutron Star with a Quark Core }
The equilibrium energy density distribution of slowly rotating
spherical star is determined by the Tolman-Oppenheimer-Volkoff
equation (TOV) \cite{shapiro1983, glendenning2000, weber1999},

\begin{equation}\label{eq12}
    \frac{dP}{dr}=-\frac{G[{\cal E}(r)+\frac{P(r)}{c^2}]
    [m(r)+\frac{4\pi r^3 P(r)}{c^2}]}{r^2
    [1-\frac{2Gm(r)}{rc^2}]},
\end{equation}

\begin{equation}\label{eq13}
    \frac{dm}{dr}=4\pi r^2{\cal E}(r).
\end{equation}
$P$ is the pressure and ${\cal E}$ is the total energy density. For
a given equation of state in the form $P({\cal E})$,  the TOV
equation yields the mass and radius of star as a function of central
energy density.

In our calculation for the hot neutron star with quark core, we use
the following equation of state: (i) Below the density $0.05 fm^3$,
we use the equation of state calculated by Baym
\cite{baym1971}. (ii) For the hadron phase, from the density of
$0.05 fm^3$ up to the density where the mixed phase is started, we
use the equation of state which is calculated in section \ref{Hadron
phase}. (iii) In the range of densities in mixed phase, we use the
equation of state calculated in section \ref{Mixed phase}. (iv) For
quark phase, we use the equation of state calculated in section
\ref{Quark Phase}.
 Using the above equations of state, we  integrate the
TOV equation numerically and determine the structure of this star. All
calculations are done for the density dependent bag constant (${\cal B}(n)$) at two
different temperatures $T=10$ and $20 MeV$. Our results are as follows.

\begin{figure}
\includegraphics[scale=0.45]{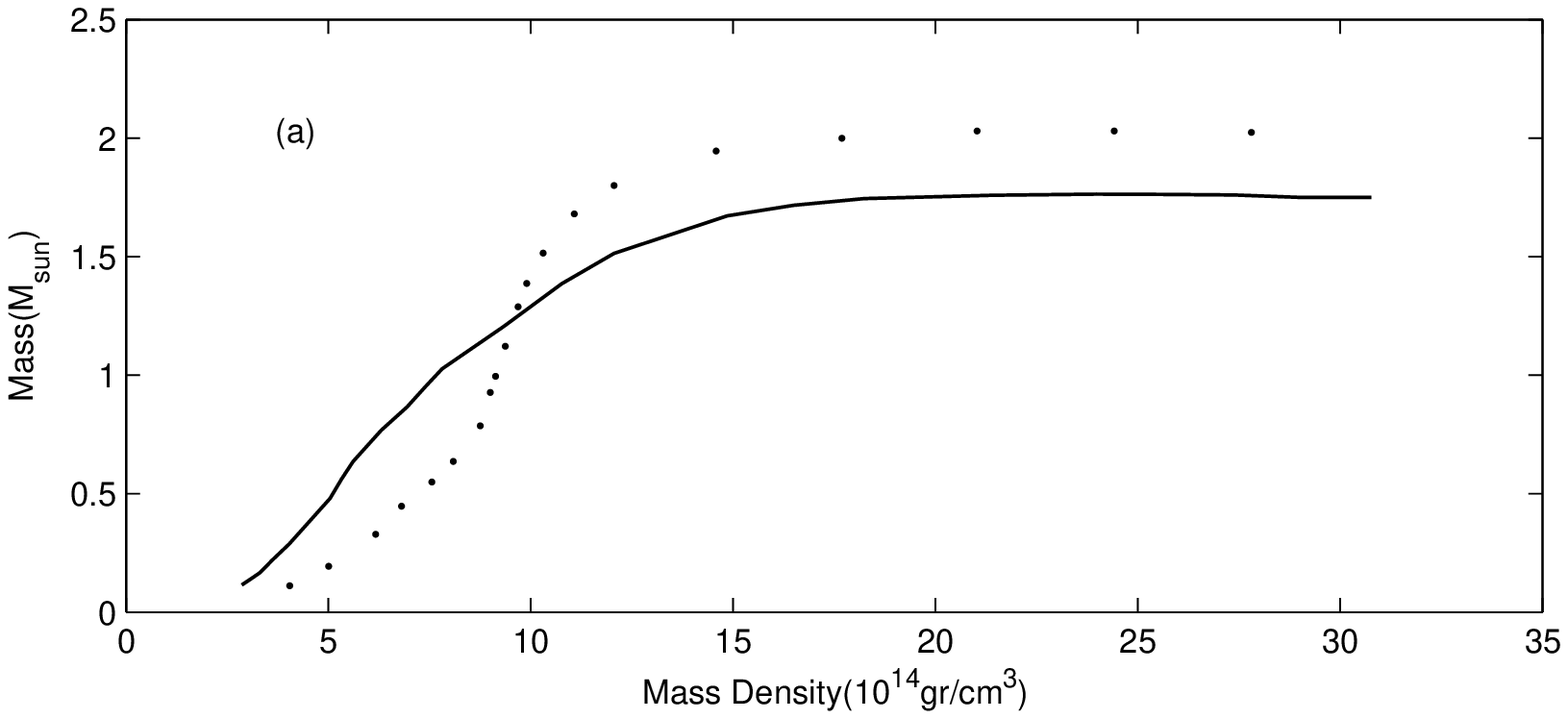}
\includegraphics[scale=0.45]{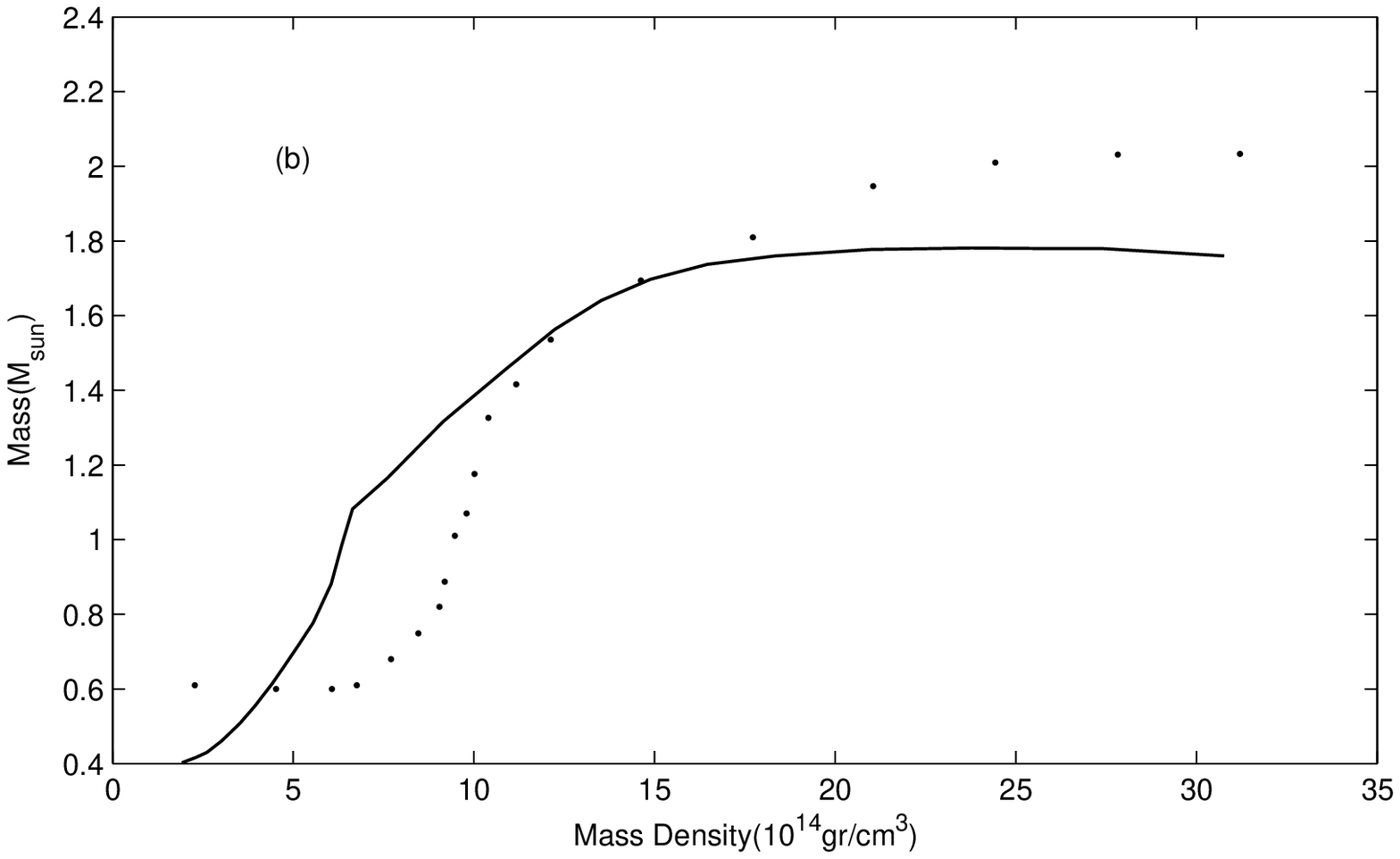}

\caption{Gravitational mass versus the central mass density for the neutron star with the quark core with
density dependent (dotted line) and independent (solid line) bag constant  at $T = 10$ (a) and $20\ MeV$ (b).}
\label{mass10}
\end{figure}

\begin{figure}

\includegraphics[scale=0.45]{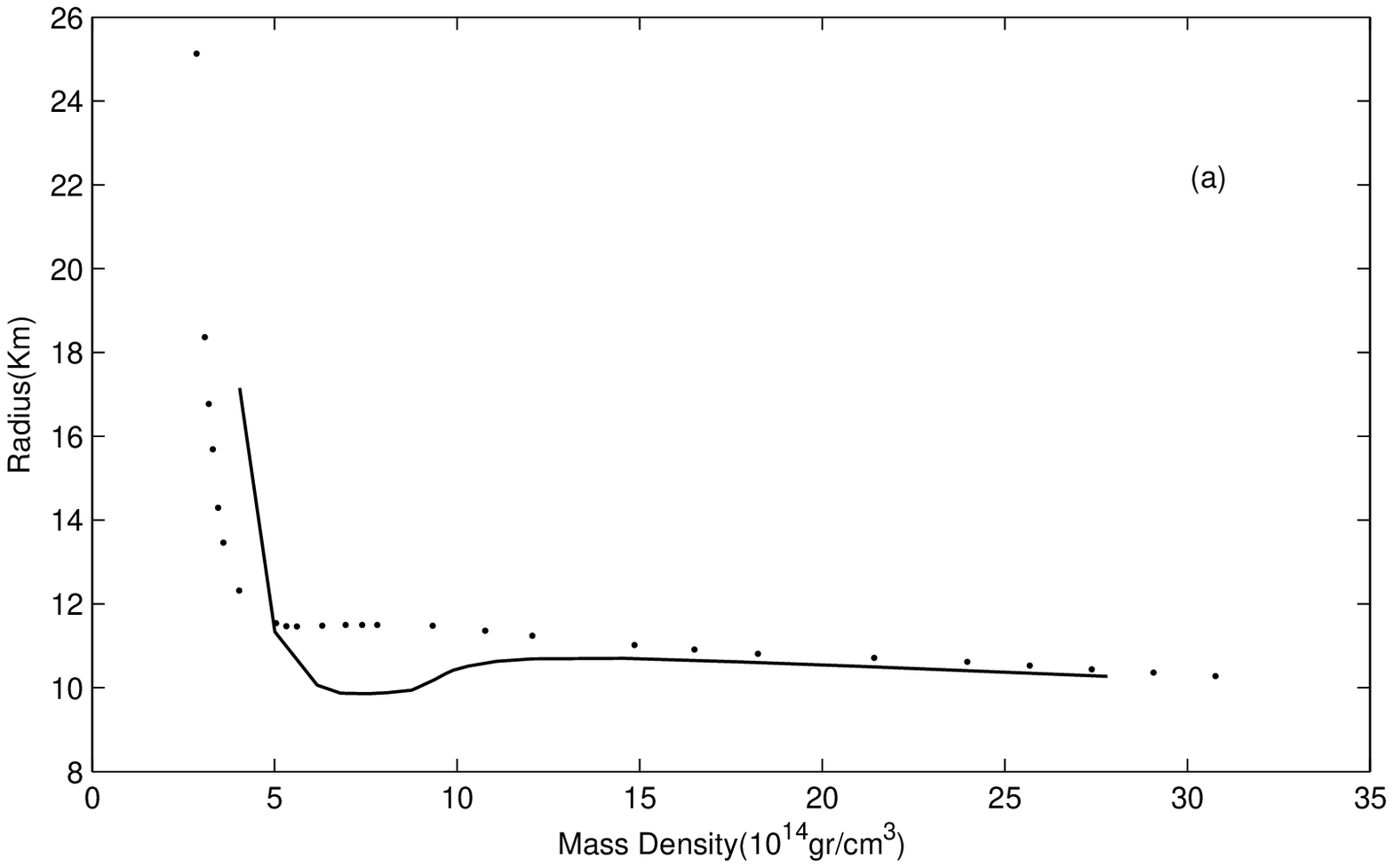}
\includegraphics[scale=0.45]{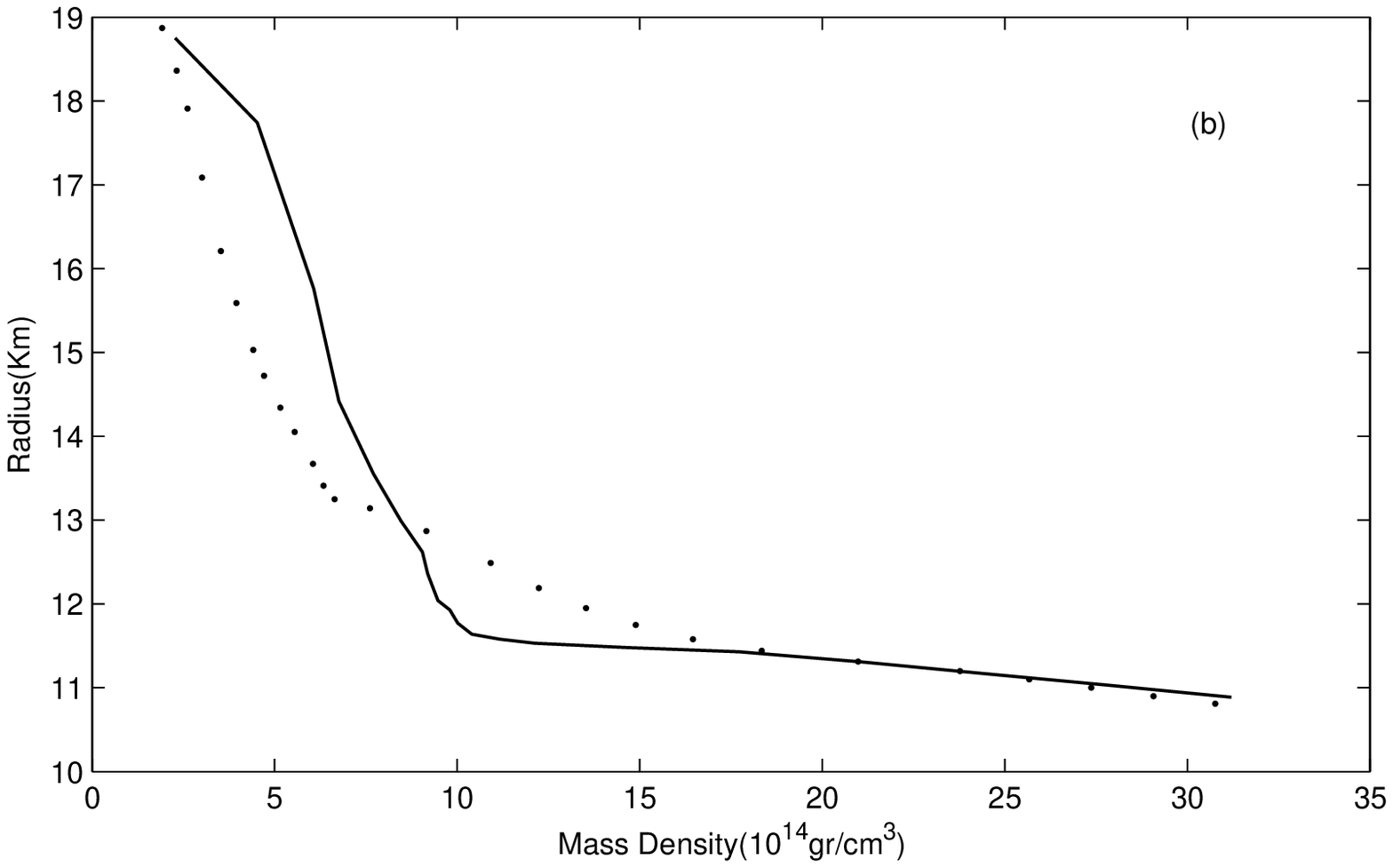}

\caption{radius versus the central mass density for the neutron star with the quark core with
density dependent (dotted line) and independent (solid line) bag constant  at $T = 10$ (a) and $20\ MeV$ (b).} \label{radius10}
\end{figure}

The gravitational mass versus the central mass density for a hot
neutron star with quark core for two different temperatures has been
presented in Fig. \ref{mass10}. We can see there is
the limiting mass for hot neutron star and  this mass increases when
we consider a density dependent bag constant in MIT model. This is
reasonable because when a density dependent bag constant is
considered, the equation of state is softer as this is seen in Fig.
\ref{pn1f,n} for two temperatures. The radius as
a function of central mass density for hot neutron star with quark
core for two different temperatures has been presented in Fig.
\ref{radius10}. The radius of star decreases when
the bag constant is density dependent. Our results for the maximum
gravitational mass of the hot neutron star with the quark core and
the corresponding values of radius and central mass density have
been given in Tables 1 and 2 for two different
temperatures.

\section{Summary and Conclusion}

The structure of hot neutron star with a quark core using a density
dependent bag constant has been investigated.
From the surface toward the center of hot neutron star, a pure
hadronic matter, a mixed phase of quarks and hadrons in a range of
densities determined by employing the Gibbs conditions, and a pure
quark matter in the core, have been considered.
We have employed the LOCV method at finite temperature to get the
equation of state of hot hadronic matter.
The MIT bag model with the density dependent bag constant obtained
by using the recent experimental results of CERN SPS on the
formation of a quark-gluon plasma has been applied to compute the
equation of state of hot quark matter.
We have solved the TOV equation by a numerical method to determine
the structural properties of hot neutron star with the quark core at
T = 10 and 20 MeV.
The results have been compared to those for the hot neutron star
with ${\cal B}=90\ MeV/fm^3$.
We have found that a density dependent bag constant leads to a
higher mass and a lower radius for the neutron star in comparison to
the case in which the constant ${\cal B}=90\ MeV/fm^3$ has been used.

\section*{Acknowledgements}
{We wish to tank the Research Council of Islamic
Azad University, Bafgh Branch. G. H. Bordbar wishes to thank the
Research Institute for Astronomy and Astrophysics of Maragha and Shiraz University Research Council for financial support.}



\begin{table}[h]
\label{tab1}
\begin{center}
  \caption[]{Maximum gravitational mass $(M_{max})$ and the corresponding
  radius $(R)$ and central mass density $({\cal E}_c)$ of hot neutron star with a quark core for
  density dependent and fixed ${\cal B}$ at $T=10\ MeV$.}
  \begin{tabular}{clclcl}
  \hline\noalign{\smallskip}
 & $NS+Quark Core$ & $M_{max}\left(M_{\,\odot}\right)$ & $R\left(km\right)$ & ${\cal E}_c\left({10^{14}gr/cm^3}\right)$ \\
 \hline\noalign{\smallskip}
 & Dependent ${\cal B}$   & 2.032 & 10.39 & 24.42  \\
 & Fixed ${\cal B}$  & 1.76 & 10.45 & 27.38  \\
 \noalign{\smallskip}\hline
  \end{tabular}
\end{center}

\end{table}
\begin{table}[h]
\label{tab2}
\begin{center}
  \caption[]{Maximum gravitational mass $(M_{max})$ and the corresponding
  radius $(R)$ and central mass density $({\cal E}_c)$ of hot neutron star with a quark core for density dependent and fixed ${\cal B}$ at $T=20\ MeV$.}

  \begin{tabular}{clclcl}
  \hline\noalign{\smallskip}
 & $NS+Quark Core$ & $M_{max}\left(M_{\,\odot}\right)$ & $R\left(km\right)$ & ${\cal E}_c\left({10^{14}gr/cm^3}\right)$ \\
 \hline\noalign{\smallskip}

 & Dependent ${\cal B}$   & 2.033 & 10.9 & 24.43  \\
 & Fixed B  & 1.78 & 11 & 27.37  \\

 \noalign{\smallskip}\hline
  \end{tabular}
\end{center}

\end{table}
\newpage

\begin{figure}

\includegraphics[scale=0.45]{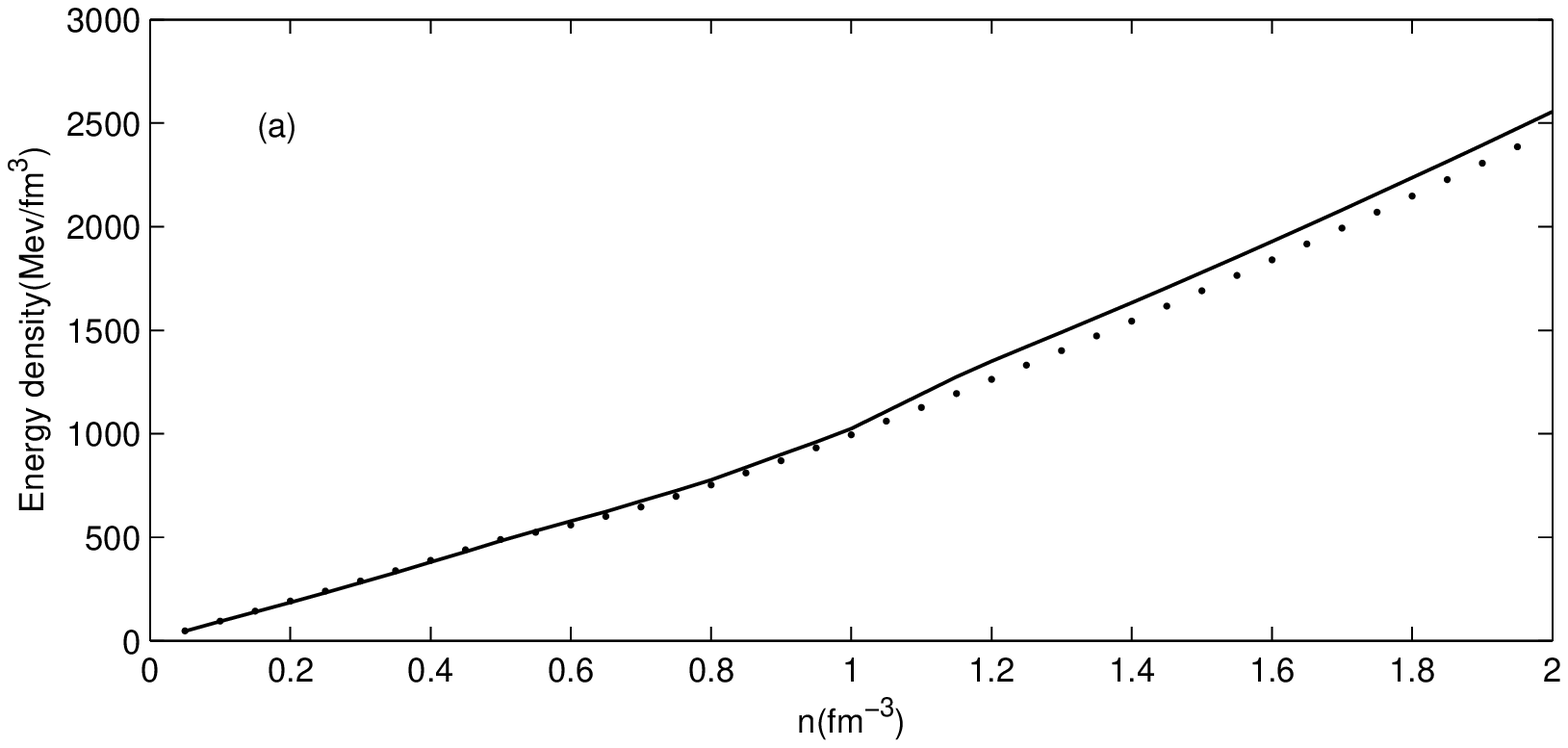}
\includegraphics[scale=0.45]{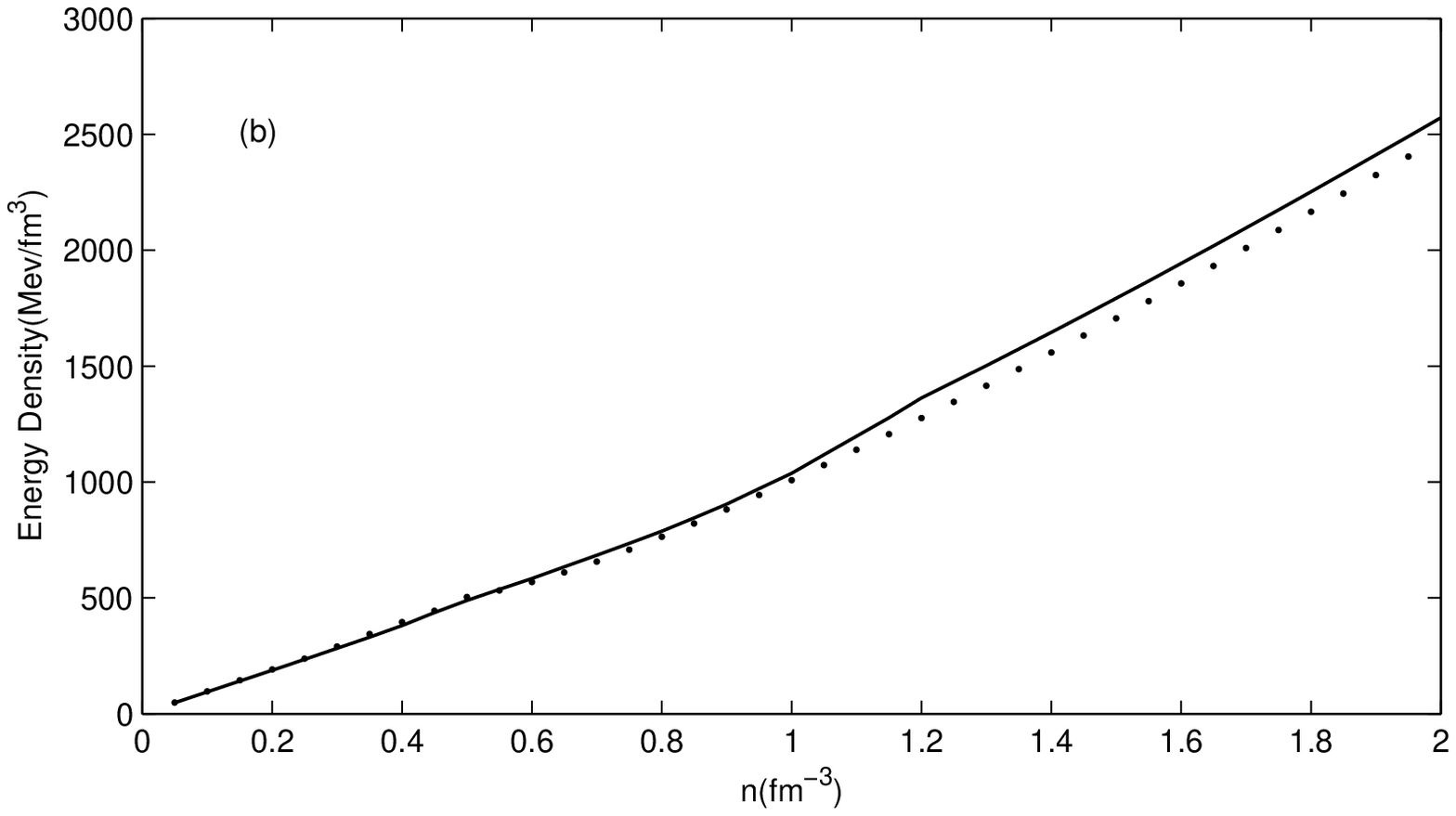}

\caption{Energy density versus the baryon density for the neutron star with the quark core with
density dependent (dotted line) and independent (solid line) bag constant  at $T = 10$ (a) and $20\ MeV$ (b).}
\label{en1f,n}
\end{figure}


\begin{figure}
\includegraphics[scale=0.45]{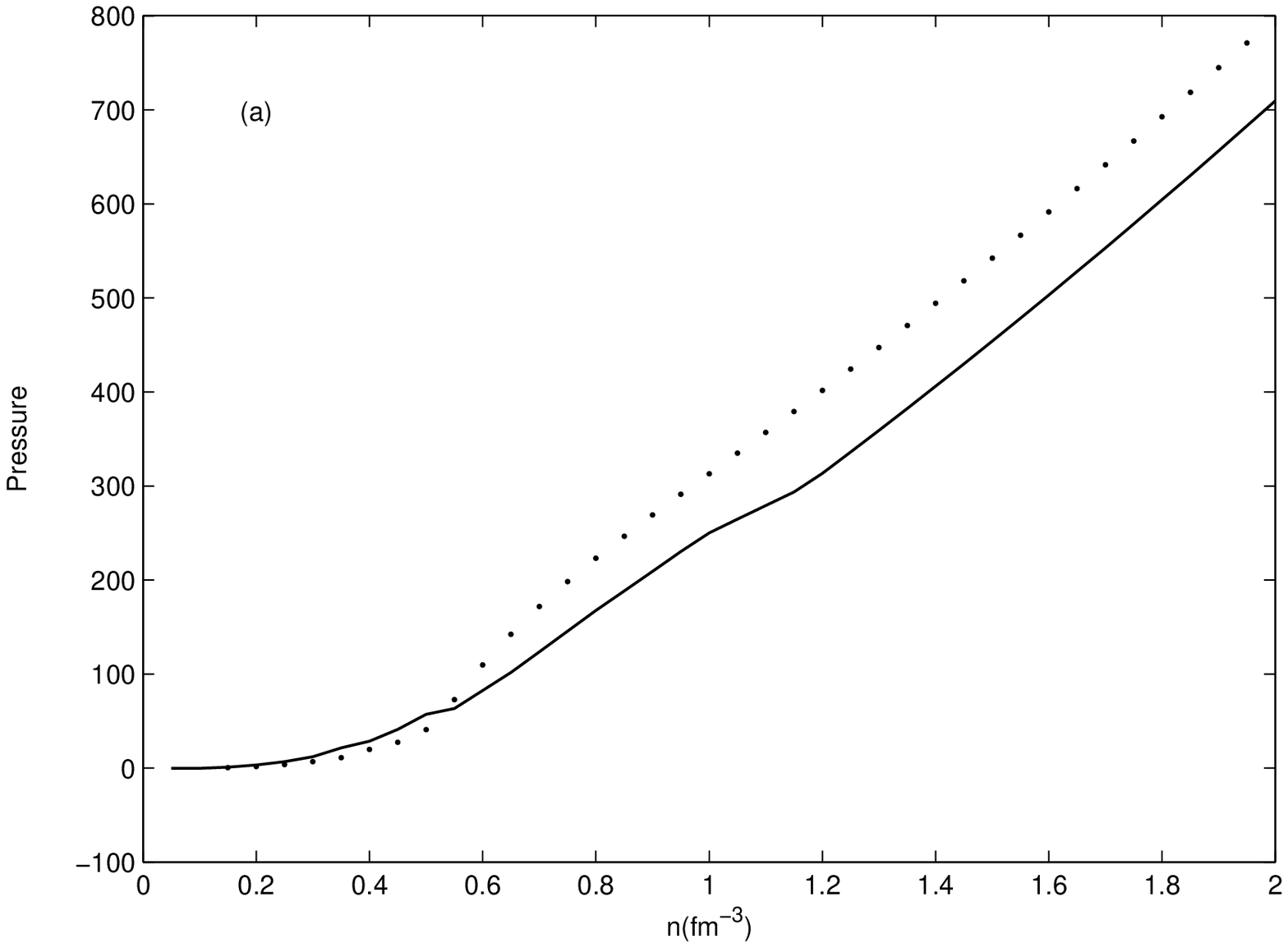}
\includegraphics[scale=0.45]{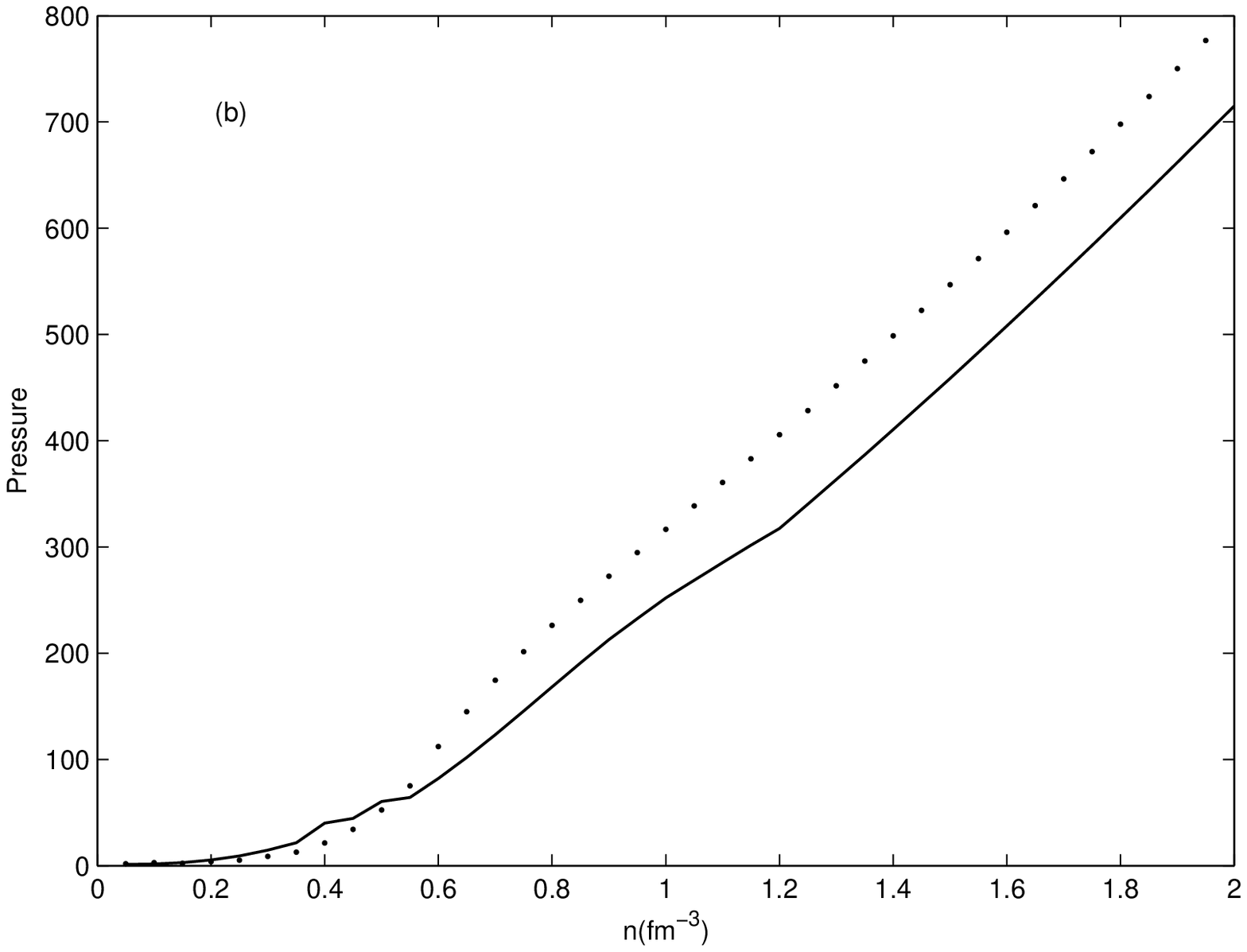}

\caption{Pressure versus the baryon density for the neutron star with the quark core with
density dependent (dotted line) and independent (solid line) bag constant  at $T = 10$ (a) and $20\ MeV$ (b).}
\label{pn1f,n}
\end{figure}

\newpage

\begin{thebibliography}{99}

\bibitem{burrows1986} A. Burrows and J. M.Lattimer,\emph{ ApJ} {\bf 307}, 178 (1986)
\bibitem{glendenning1992} N. K. Glendenning, \emph{ Phys. Rev.} {\bf D46}, 1274 (1992)
\bibitem{chodos1974}A. Chodos, R.L. Jaffe, K. Johnson, C. B. Thorn and V. F. Weisskopf, \emph{Phys. Rev.} {\bf D9}, 3471( 1974).
\bibitem{rafelsky1982} J. Rafelsky and B. Muller, \emph{ Phys. Rev. Lett.} {\bf 48} 1066(1982)
\bibitem {matsui1986} T. Matsuiand  and H.Satz, \emph{Phys. Lett.} {\bf B178}, 416(1986)
\bibitem {cleman1986}J. Cleymans, R.V. Gavai and E. Suhonen, \emph{ Physics Rep.} {\bf 130}, 217 (1986).
\bibitem {Burgio2002a} G. F. Burgio, M. Baldo, P. K. Sahu and H.-J. Schulze, \emph{ Phys. Rev.} {\bf C66}, 025802(2002)
\bibitem{burgio2007} G. F. Burgio, M. Baldo, O. E. Nicotra and H.-J. Schulze,\emph{ Ap\&SS }{\bf 308}, 387(2007)
\bibitem{bordbar2006} G. H. Bordbar, M. Bigdeli and T. Yazdizadeh, \emph{Int. J. Mod. Phys.} {\bf A21}, 5991 (2006)
\bibitem{bordbarh2006} G. H. Bordbar and M. Hayati, \emph{ Int. J. Mod. Phys.} {\bf A21}, 1555 (2006)
\bibitem{yazdizadeh2011}  T. Yazdizadeh and G.H. Bordbar, \emph{ Res. Astron. Astrophys} {\bf 11}, 471(2011)
\bibitem{bordbar1997}  G. H. Bordbar and M. Modarres, \emph{ J. Phys. G: Nucl. Phys.} {\bf 23}, 1631(1997)
\bibitem{bordbar1998} G. H. Bordbar and M. Modarres, \emph{ Phys. Rev.} {\bf C57}, 714.(1998)
\bibitem{modarres1998} M. Modarres and G. H. Bordbar, \emph{ Phys. Rev.} {\bf C58}, 2781(1998)
\bibitem{bordbar2007a}G. H. Bordbar and M. Bigdeli, \emph{ Phys. Rev.} {\bf C75}, 045804(2007)
\bibitem{bordbar2007b}G. H. Bordbar and M. Bigdeli, \emph{ Phys. Rev.} {\bf C76}, 035803(2007)
\bibitem{bordbar2008a} G. H.Bordbar and Bigdeli \emph{ Phys. Rev.} {\bf C77}, 015805(2008)
\bibitem{bordbar2008b} G. H.Bordbar and Bigdeli \emph{ Phys. Rev.} {\bf C78}, 054315(2008)
\bibitem{bigdeli2009} M. Bigdeli, G. H. Bordbar and Z. Rezaei,\emph{ Phys. Rev.} {\bf C80}, 034310 (2009).
\bibitem{farhi1984}  E.Farhi, R. L. Jaffe, \emph{ Phys. Rev.} {\bf D30}, 2379(1984)
\bibitem{jacob2000}  U.Heinz and M. Jacobs, nucl-th/0002042.
\bibitem{heinz2000}  U.Heinz, \emph{ Nuch. phys.} {\bf A685}, 414(2001)
\bibitem{adami1993} C. Adami and  G.E. Brown, \emph{ Phys. Rep.} {\bf 234}, 1(1993)
\bibitem{blaschke1999} D. Blaschke, H. Grigorian, G. Poghosyan, C.D. Roberts and S. Schmidt, \emph{ Phys. Lett.
} {\bf B450}, 207(1999)
\bibitem {Baldo2006}  M. Baldo, G. F. Burgio and H.-J. Schulze, \emph{SUPERDENSE QCD MATTER AND COMPACT STARS NATO Science Series}{\bf Volume
197, Part II}, 113 (2006)
\bibitem {Burgio2002b}G. F. Burgio, M. Baldo ,P. K. Sahu, A. B. Santra and H.-J. Schulze,
\emph{Phys. Lett.} {\bf B526}, 19 (2002)
\bibitem{shapiro1983} S. L. Shapiro and S. A. Teukolski, 1983, \emph{Black Holes,
White Dwarfs and neutron Stars} (New York ).
\bibitem{glendenning2000} N. K. Glendenning, \emph{ Compact Star, Nuclear Physics, Particle Physics, and
General Relativity} (Springer, New York), 2nd  ed.(2000)
\bibitem{weber1999} F. Weber, \emph{Pulsars as Astrophysical Laboratories for Nuclear and
Particle Physics} (Institute of Physics, Bristol)(1999)
\bibitem{baym1971} G. Baym, C. Pethick and P. Sutherland, \emph{ Astrophys. J.} {\bf 170}, 299(1971)

\end{thebibliography}
\end{document}